# Theoretical investigation of a CDDW Hamiltonian for YBCO to explore the possibility of quantum oscillations in specific heat

**Partha Goswami[1]**

[1]D.B.College, University of Delhi, Kalkaji, New Delhi-110019, India

**PACS** 74.20.-z .

**Abstract.** We investigate a chiral d-density wave (CDDW) mean field Hamiltonian in momentum space for the under-doped YBCO. We consider the commensurate density wave order which doubles the unit cell of the real space lattice and derive an expression for entropy in the CDDW state in the absence of magnetic field. We show that the transition to CDDW state gives rise to the occurrence of discontinuities in the entropy density difference between the CDDW state and the normal state at eight points(Hot spots) on the boundary of the reduced Brillouin zone. The chirality induced anomalous Nernst signal is shown to arise mainly from the points close to these Hot spots and not from the hole pockets as reported by the previous authors. We also show the possibility of quantum oscillations in the specific heat for the system under consideration in the presence of a changing magnetic field.

**1 Introduction** In this communication, starting with a chiral d-density wave (CDDW)[1] mean field Hamiltonian in momentum space for the pseudo-gapped state of under-doped YBCO, we show that the approximate (1/B)-oscillations in specific heat is possible. The two main frequencies of the oscillations have their origin in the electron and the hole pockets of the reconstructed Fermi-Luttinger surface obtained using the 'maximal gradient method'[2]. From the thermodynamics viewpoint an important point of distinction between these pockets is that the momentum dependent pressure corresponding to the former is positive whereas that corresponding to the latter is negative. We also report that at certain points ('Hot spots') on the boundary of the reduced Brillouin zone(RBZ) close to the Fermi pockets there are discontinuities in the entropy density difference (EDD) between the CDDW state and the normal state; on the boundaries of the electron and hole pockets EDD peak dramatically (see Fig.1). We find that the main contribution to the chirality induced anomalous Nernst signal (ANS) comes from the points close to these 'Hot spots' and not from the hole pockets in the momentum space as reported by the previous workers[3]. The calculation of the momentum dependent anomalous dc Hall conductivity (AdcHC), however, has confirmed that the corresponding peaks are located at the hole-pockets which is in agreement with ref.[3]. The paper is organized as follows: In section 2 we outline the derivations of the entropy, and the chirality driven coefficient of the transverse heat current. We conclude the paper in section 3 describing the theoretical investigation of the quantum oscillation in specific heat, in the presence of a changing magnetic field.

## 2 Entropy density and anomalous Nernst effect

**2.1 Entropy density** In the second-quantized notation, the Hamiltonian (with index i = (1,2) below corresponding to two layers of YBCO) for the Chiral d+id density-wave state can be expressed as $H_{d+id} = \sum_{k,\sigma,i=1,2} \Phi^{(i)\dagger}_{k,\sigma} E(k) \Phi^{(i)}_{k,\sigma}$ where $\Phi^{(i)\dagger}_{k,\sigma} = (d^{\dagger(1)}_{k,\sigma}\ d^{\dagger(1)}_{k+Q,\sigma}\ d^{\dagger(2)}_{k,\sigma}\ d^{\dagger(2)}_{k+Q,\sigma})$ and $E(k) = [\varepsilon_k^U I_{4\times 4} + \zeta_k \cdot \alpha]$. Here $\alpha = (\alpha_1\ \alpha_2\ \alpha_3\ \alpha_4)$ with

$$\alpha_i = \begin{pmatrix} \sigma_i & 0 \\ 0 & \sigma_i \end{pmatrix}\ (i=1,2,3),\ \alpha_4 = \begin{pmatrix} 0 & I_{2\times 2} \\ I_{2\times 2} & 0 \end{pmatrix}. \quad (1)$$

$I_{4\times 4}$ and $I_{2\times 2}$, respectively, are the 4×4 and 2×2 unit matrices; $\sigma_i$ are the Pauli matrices and $\zeta_k = (-\chi_k\ -\Delta_k\ \varepsilon_k^L\ t_k)$ – a four-component vector. Here the chiral order parameter[3], $D_k \exp(i\theta_k)$, is given by $D_k = (\chi_k^2 + \Delta_k^2)^{1/2}$ and $\cot\theta_k = (-\chi_k/\Delta_k)$ with $\chi_k = -(\chi_0/2)\sin(k_x a)\sin(k_y a)$, and $\Delta_k = (\Delta_0(T)/2)(\cos k_x a - \cos k_y a)$. The quantity $\varepsilon_k$ is the usual [4] normal state tight-binding energy dispersion involving t, t´, t´´ which are the hopping elements between the nearest, next-nearest(NN) and NNN neighbors, respectively; $\mathbf{Q} = (\pm\pi, \pm\pi)$ is the ordering vector. The quantity $t_k$ is momentum conserving tunneling matrix element which for the tetragonal structure is given by $t_k = (t_0/4)(\cos k_x a - \cos k_y a)^2$. The energy eigenvalues of E(k) (with j = ± 1 ( j = +1 corresponds to the upper branch(U) and j = –1 to the lower branch(L) ), and ν = ± 1) are $E_k^{(j,\nu)} = [\varepsilon_k^U - \mu + jw_k + \nu t_k]$ where $\varepsilon_k^U = (\varepsilon_k + \varepsilon_{k+Q})/2$, $\varepsilon_k^L = (\varepsilon_k - \varepsilon_{k+Q})/2$, $w_k = [(\varepsilon_k^L)^2 + D_k^2]^{1/2}$, and μ is the chemical potential. With these eigenvalues, we find that the momentum dependent occupancy (MDO)

$$f(k) = \sum_{\nu,\sigma} [u_k^{(\nu)2} (\exp(\beta E_k^{(U,\nu)}) + 1)^{-1} + v_k^{(\nu)2} (\exp(\beta E_k^{(L,\nu)}) + 1)^{-1}] \quad (2)$$

and the density of states (DOS) $\rho(k,\omega) = \sum_{\nu,\sigma}[u_k^{(\nu)2}\delta(\omega - E_k^{(U,\nu)}) + v_k^{(\nu)2}\delta(\omega - E_k^{(L,\nu)})]$ where $u_k^{(\nu)2} = (1/4)[1 + (\varepsilon_k^L/w_k)]$ and $v_k^{(\nu)2} = (1/4)[1 - (\varepsilon_k^L/w_k)]$.

The thermodynamic potential is given by

$$\Omega = \Omega_0 - 2(\beta N_s)^{-1} \sum_{k,r(=U,L),\nu} \{\ln\cosh(\beta E_k^{(r,\nu)}/2)\} \quad (3)$$

where $\Omega_0 = (N_s)^{-1}\sum_{k,r(=U,L),\nu} E_k^{(r,\nu)}$, $\beta = (k_B T)^{-1}$ and $N_s$ is the number of unit cells. We obtain the Fermi-Luttinger surface (FLS), with small electron and hole pockets, using the expression for f(k) and applying the 'maximal gradient method'[2]. The method is based on the fact that the FLS is given by the set of **k**-values for which the MDO, f(k), shows a jump discontinuity in the zero temperature limit. When this discontinuity is smeared out, say, by thermal broadening, the gradient of f(k), |∇f(k)|, is assumed to be maximal at the locus of the underlying FLS. The electron pockets correspond to the (anti-nodal) sector centred around [(±π, 0), (0,±π)] where $\chi_k \to 0$. On the other hand, the hole pockets are located near the (nodal) region centered around (±π/2, ±π/2) where the gap $\Delta_k$ is zero. The dimensionless entropy per unit cell is given by $S = \beta^2 (\partial\Omega/\partial\beta)$. We obtain for the pseudo-gapped(PG) phase and the normal phase $S_{PG} = (2/N_s)\sum_{k,r(=U,L),\nu} s_{PG}(\mathbf{k})$ and $S_N = (2/N_s)\sum_{k,j=1,2} s_N(\mathbf{k})$ where

$$s_{PG}(\mathbf{k}) = [\ln(1/2) + \ln(1 + \exp(-\beta E_k^{(r,\nu)})) + (\beta E_k^{(r,\nu)} + \beta^2 (\partial E_k^{(r,\nu)}/\partial\beta))(\exp(\beta E_k^{(r,\nu)}) + 1)^{-1}], \quad (4)$$

$$s_N(\mathbf{k}) = [\ln(1/2) + \ln(1 + \exp(-\beta \varepsilon_j(k))) + (\beta\varepsilon_j(k) + \beta^2 (\partial\varepsilon_j(k)/\partial\beta))(\exp(\beta\varepsilon_j(k)) + 1)^{-1}], \quad (5)$$

$\varepsilon_1(k) = \varepsilon_k$, and $\varepsilon_2(k) = \varepsilon_{k+Q}$. The parameters we choose (see ref.[4]) for the calculation of EDD, ANS, specific heat, etc. at 10% hole doping, are: t = 0.25 eV, t´ = 0.4t, t´´ = 0.0444 t, μ = –0.27 eV, and $t_0 = 0.032$ t. The pseudo-gap(PG) temperature T* ~ 155 K. We assume, the experimental value $\Delta_0$ ( T < T*) = 0.0825 eV = 0.3300 t in the vicinity of T*, and $(\chi_0/\Delta_0 (T<T^*))^2 = 0.0025$. The expression for the thermodynamic potential above yields the free energy density (at a temperature T = 150 K) $\beta\Omega$ (**k**=[(±π,0), (0,±π)]) = –102.2992 and $\beta\Omega$ (**k**= (±π/2,±π/2)) = 5.5452. Thus the fermion pressure at the anti-nodal point is positive while that at the nodal point is negative. We find that at the 'Hot spots' (location ~ (±0.64, ±2.50), ( ±2.50, ±0.64)) on the boundary of RBZ there are discontinuities in $\Delta s(\mathbf{k}) = (s_{PG}(\mathbf{k}) - s_N(\mathbf{k}))$; on the boundaries of the Fermi pockets $\Delta s(\mathbf{k})$ peak dramatically.

**Table1.** The data in the table indicates that the in $\Delta s(\mathbf{k})$ occurs at the point ~(2.50,0.64) on the boundary of the reduced Brillouin zone.

| $k_x a$ | 2.41 | 2.50 | 2.60 | 2.65 |
|---|---|---|---|---|
| $k_y a$ | 0.73 | 0.64 | 0.54 | 0.49 |
| $\Delta s(\mathbf{k})$ | 0.7630 | –1.2109 | 0.5405 | 1.2281 |

The contour plot of the entropy density difference on the Brillouin zone at 10% hole doping is shown in Fig. 1. Here the small, deep blue squares on the RBZ boundary near the Fermi pockets are the so-called 'Hot spots'. The $\Delta s(\mathbf{k})$ peaks are shown by the bright orange outlines. The entropy difference $\sum \Delta s(\mathbf{k})$ is found to be positive, and therefore the transition to pseudo-gapped phase is first order.

**2.2 Anomalous Nernst signal** In the presence of an external electric field **E** along y-direction (and magnetic field **B** =0), the transverse heat current $J_x$ in the x-direction is given by the relation $J_x = (T S_{xy} E_y)$ where the coefficient $S_{xy}$ is defined below. The reason for nonzero $J_x$ is that, in the presence of chirality, the carriers acquire an anomalous velocity $\mathbf{v}_a$ given by $\hbar \mathbf{v}_a = e \mathbf{E} \times \mathbf{\Omega}^{(a)}(\mathbf{k})$ where $\Omega^{(a=1,2)}(k_x, k_y) = \pm ta^2 w_k^{-3} \chi_0 \Delta_0 (\sin^2 k_y a + \sin^2 k_x a \cos^2 k_y a)$ are the Berry curvatures having opposite signs with nonzero component only in the z-direction (see Ref.[3]). Upon multiplying this velocity by the entropy density we obtain the coefficient $S_{xy}$ for the transverse heat current: $S_{xy}(T<T^*) = (e/\hbar)(1/(N_s a^2))\sum_k \Omega^{(a)}(\mathbf{k}) k_B s_{PG}(\mathbf{k})$ where $\Omega^{(1)}(\mathbf{k})$ (> 0) is multiplied with $s_{PG}(\mathbf{k})$ corresponding to the upper branch (U) and $\Omega^{(2)}(\mathbf{k})$(< 0) to that for the lower branch(L). Upon taking $(\chi_0/\Delta_0 (T<T^*))^2 = 0.0025$, we find graphically, as well as numerically, that $N(k) = (a^{-2}\sum_a \Omega^{(a)}(\mathbf{k}) s_{PG}(\mathbf{k}))$ peaks at the points ((±0.54, ±2.60),( ±2.60, ±0.54)), very close to the aforementioned 'Hot Spots', on the boundary of RBZ with peak value ~ 1.8. Elsewhere on the BZ, including the hole pockets, N(k) is practically zero (see Table 2). The calculation of the momentum dependent AdcHC[3], Σ(**k**), on the other hand shows that its origin lies in the hole pockets (see Table 2).

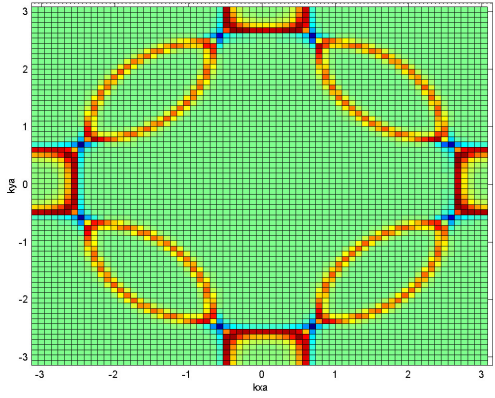

**Figure 1** The contour plot of EDD on the Brillouin zone at 10% hole doping is shown above. Here the deep blue squares on the RBZ boundary and near the Fermi pockets are the so-called 'Hot spots'.

**Table2.** The data in the table indicates that the peak in N(**k**) occurs at (2.60,0.54) and not at the nodal point (1.57,1.57). The table also indicates that the anomalous dc Hall conductivity corresponds to a hole-pocket-based activity.

| $k_x a$ | 2.55 | 2.60 | 2.65 | 1.57 |
|---|---|---|---|---|
| $k_y a$ | 0.59 | 0.54 | 0.49 | 1.57 |
| N(k) | 1.6793 | 1.7614 | 1.6846 | –0.0000 |
| $\Sigma(\mathbf{k})$ | –0.0000 | –0.0000 | –0.0000 | 2.0449 |

We find that the anomalous Nernst signal measure $S_{xy}\rho$ ~ 95 nV-K where $\rho$ is the magneto-resistivity ~ 0.3 ohm-m for YBCO. In high-temperature superconductors, such as LSCO and BSCCO at the temperature close to T*, the experimentally observed ANS ~ 100 nV-K$^{-1}$.

**3. Quantum oscillations in specific heat** We shall now discuss the specific heat oscillation at low temperature in the presence of a magnetic field. For a magnetic field applied in z-direction ( i.e. the vector potential **A**=(0,−Bx, 0) in Landau gauge),starting with a dispersion $\varepsilon_k(B)$ where the first term includes the chemical potential $\mu$, the Landau levels and cyclotron frequenc $\omega_c = eB/m^*$ ($m^*$ is the effective mass of the electrons), and the Zeeman term (g $\mu_B B/2$), viz. $\varepsilon_k(B) = -\mu' - 2t(\cos(k_x a) + \cos(k_y a + \varphi)) + 4t' \cos(k_x a) \cos(k_y a + \varphi/2)$, $\mu' \equiv \mu - \hbar \sum_{n=0}^{\infty} (2n+1)(\omega_c/2) + (-1)^\sigma (g\mu_B B/2)$, n = 0, 1...(The quantity $\varphi = (2\pi e B a^2/h)$ is the Peierls phase factor and 'a' is the lattice constant of YBCO), we find that the specific heat (C = $-\beta (\partial S_{PG}/\partial\beta)$) at a given doping level is given by $C \approx (k_B^2 T/2\pi t) \int_{-\infty}^{+\infty} dx\, I(x,B)\{x^2 e^x/(e^x+1)^2\}$ where $I(x,B)=\int d\mathbf{k} \sum_{v,\sigma}[u^{(v)}_k{}^2 \delta((x/\beta t)-(E'^{(U,v)}_k/t)) + v^{(v)}_k{}^2 \delta((x/\beta t)-(E'^{(L,v)}_k/t))]$, $E'^{(j,v)}_k = E^{(j,v)}_k + \hbar \sum_{n=0}^\infty (2n+1)(\omega_c/2)$, and $\int d\mathbf{k} \to \int_{-\pi}^{+\pi}(d(k_x a)/2\pi \int_{-\pi}^{+\pi}(d(k_y a)/2\pi$ ; the Zeeman term is assumed to be insignificant. In the absence of magnetic field (B), the bunch of delta functions in the integrand in I(x,B) can be expressed in terms of Lorentzians by taking into consideration the scattering by impurities. The specific heat will have the usual linear dependence on temperature (T). The present problem, however, is to calculate specific heat when B $\neq$ 0. We shall not consider the scattering by impurities aspect in this communication. The delta functions can now be expanded in cosine Fourier series: $\delta(x-a) = (2\pi)^{-1} + \pi^{-1}\sum_{m=1}^\infty \cos[m(x-a)]$. Upon doing so, we find that the non-oscillatory part of the specific heat, with the linear dependence on (T/B), is $C_{non-oscll} \sim (2k_B^2 T/3\hbar\omega_c)$. The oscillatory part, for B $\neq$0, may be expressed as $C_{oscll} = (2k_B^2 T/\pi^2 \hbar\omega_c)\int d\mathbf{k}\, q(B,\mathbf{k})$ where

$$q(B,\mathbf{k})=\int_{-\infty}^{+\infty}dx\sum_{m=1}^\infty \cos(2mx/\beta\hbar\omega_c)\{x^2 e^x/(e^x+1)^2\}$$
$$\times \sum_{v,\sigma}[u^{(v)}_k{}^2 \cos\{m(|2E_k^{(U,v)}/(\hbar\omega_c)|-\theta)\}$$
$$+v^{(v)}_k{}^2 \cos\{m((2E_k^{(L,v)}/\hbar\omega_c)+\theta)\}] \quad (6)$$

and $\theta = \sum_{n=0}^\infty (2n+1)$. Equation (6) indicates that the origin, of the approximate (1/B)-oscillations in specific heat, is the upper and the lower branches of the excitation spectrum (or the electron and the hole pockets, respectively). In order to estimate the oscillation frequencies($F_U$ and $F_L$), we make the approximation $\{E_k^{(U,v)} \approx E_k^{(U,v)}|_{\mathbf{k}=[(\pm\pi,0),(0,\pm\pi)]} \equiv \bar{E}_U$, $E_k^{(L,v)} \approx E_k^{(L,v)}|_{\mathbf{k}=(\pm\pi/2,\pm\pi/2)} \equiv \bar{E}_L\}$ which simplifies our task greatly. We find, for the anti-nodal region (centre: $[(\pm\pi,0),(0,\pm\pi)]$), the frequency is $F_U$ ~ 230 T whereas, for the nodal sector (centre: $(\pm\pi/2,\pm\pi/2)$), $F_L$ ~ 1200 T. Using these values and the Onsager relation [5] we also find that, whereas electron pockets cover an area approximately 2.9% of the first Brillouin zone (BZ), the hole pockets cover an area approximately 5.2 times bigger. The key input in the calculation above is the form of the dispersion $\varepsilon_k(B)$; the chirality aspect plays a minor role as these oscillations are also possible in the pure d-density wave state[4]. These periodic quantum oscillations are in principle observable in all solid state properties of YBCO in the pseudo-gapped state, and have in fact been observed in conductivity (Shubnikov-de Haas oscillations) and magnetic susceptibility( de Haas-van Alphen effect) measurements recently[6,7].

In conclusion, we note that the computation of the correction to the quantum oscillations due to the Berry phase is an important task. We hope that it is likely to offer a theoretical indication of the chiral d-density wave state in the cuprates to corroborate theoretically the non-zero polar Kerr effect observed recently in YBCO [8].